\documentclass[10pt]{iopart}

\usepackage{iopams}
\usepackage{graphicx}
\usepackage[english]{babel}
\begin{document}

\title[submitted to J. Phys. Cond. Mat.]{Supercooling and freezing processes in nanoconfined water by time-resolved optical Kerr effect spectroscopy}

\author{A. Taschin$^{1}$, P. Bartolini$^{1}$, A. Marcelli$^{1}$, R. Righini$^{1,2}$ and R. Torre$^{1,3}$}

\address{ $^1$European Laboratory  for Non-Linear Spectroscopy (LENS), Univ. di Firenze, Via N. Carrara 1, I-50019 Sesto Fiorentino, Firenze, Italy.}
\address{ $^2$Dip. di Chimica, Univ. di Firenze, via Della Lastruccia 13, I-50019 Sesto Fiorentino, Firenze, Italy.}
\address{ $^3$Dip. di Fisica e Astronomia, Univ. di Firenze,
via Sansone 1, I-50019 Sesto Fiorentino, Firenze, Italy.}

\ead{torre@lens.unifi.it}

\begin{abstract}
Using heterodyne-detected optical Kerr effect (HD-OKE) measurements, we investigate the vibrational dynamics and the structural relaxation of water nanoconfined in Vycor porous silica samples (pore size $\simeq$ 4 nm) at different levels of hydration and temperatures.  At low level of hydration, corresponding to two complete superficial water layers, no freezing occurs and water remains mobile at all the investigated temperatures with dynamic features similar, but not equal, to the bulk water. The fully hydrated sample shows formation of ice at about 248 K, this process does not involve all the contained water; a part of it remains in a supercooled phase. The structural relaxation times measured from the decay of the time-dependent HD-OKE signal shows temperature dependence largely affected by the hydration level; the low frequency ($\nu <$ 500~cm$^{-1}$) vibrational spectra, obtained by the Fourier transforms of HD-OKE signal, appears less affected by confinement.
\end{abstract}


\section{Introduction}
 
The characterization and understanding of confined and interfacial water is relevant to many technological and natural processes that span from biological science to geological topics \cite{brovchenko_08}. Despite the numerous researches and investigations, several fundamental problems remain to be clarify. In fact, if liquid water interacts with a material surface the water layers at the interface show structural and dynamic alterations that turn into non-trivial modification of the its fundamental chemical-physical properties. These phenomena are particularly relevant when water is completely confined in a matrix at nanometric scale. 

Many recent physic researches focus on the investigation and measurements of water confined in porous glasses that can be considered prototype materials for these studies \cite{jahnert_08,gallo_10,molinero_10,oguni_11,lecaer_11,milischuk_11,giovambattista_12,limmer_12,milischuk_12,bertrand_13}. Among them Vycor glasses are maybe the more utilized because of their peculiar characteristics particularly useful in experimental investigations \cite{bruni_98,bellissent_98,dore_00,bellissent_01,tombari_05,tombari_05b}. These glass matrices are characterized by well defined pore diameters of nanometric dimensions, providing a tight hydrophilic confinement. In a series of previous works, we studied liquid filled Vycor samples by transient grating spectroscopy measuring the acoustic propagation \cite{cucini_07,taschin_07,taschin_08}, the liquid flow and thermal diffusion phenomena \cite{cucini_10,cucini_10b,taschin_10}. Recently, we investigated by heterodyne-detected optical Kerr effect (HD-OKE) the water dynamics in Vycor at variable hydration \cite{taschin_13b}.

Several basic questions about the physic of water confined at nanometric scale remain still open. In particular, our last study \cite{taschin_13b} provides an experimental support to the intuitive scenario that nanoconfined water can be divided in two types: the \textit{outer water} (i.e. the water layers close to pore surface) and the \textit{inner water} (i.e. the internal water layers). The outer layers present a liquid structure and dynamics modified by the interactions with the pore surfaces; the inner water seems to have characteristics very similar to the bulk water. 
The HD-OKE data at ambient temperature identify the outer water as the two water layers close to the pore surface, whereas the remaining internal 4-5 layers can be ascribed as the inner water. Here we extend the HD-OKE investigation to the lower temperatures in order to study the variations induced by the temperature to the nanoconfined water dynamics.

\section{Optical Kerr effect experiments}

Optical Kerr effect (OKE) experiments are based on impulsive generation of birefringence in an optically transparent medium induced by a linear-polarized short laser pulse (the pump) \cite{hunt_07,bartolini_08}. The dynamic processes following the impulsive excitation are measured by the polarization variations produced in a second laser pulse (the probe) which is spatially superimposed with the pump pulse into the sample. The time behaviour of the induced birefringence is reconstructed by changing the time delay between the pump and probe. The OKE signal measures the time-derivative of the correlation function of the anisotropic liquid susceptibility and it contains the relaxation processes and the vibrational response of the sample. Employing femtosecond laser pulses this experiment enables to measure the fast relaxation processes and the low frequency vibrational correlation functions of the investigated system. 

OKE investigation is able to measure dynamic regimes in a wide time window (from tens of femtoseconds to hundreds of picoseconds). So it has been useful to study several liquid matter phases; from simple molecular liquids \cite{ricci_95,bartolini_99} to liquid crystals \cite{torre_95,torre_98b}, from supercooled liquids \cite{ricci_04,torre_04,taschin_13,taschin_14} to glass-formers \cite{torre_98,torre_00,ricci_02,prevosto_02,pratesi_03}. 

The OKE signal is directly proportional to the material response function convoluted with the instrumental function, $G(t)$\cite{bartolini_08,taschin_14}:
\begin{equation}
S(t)=\int \left[ k\delta(t-t')+R_n(t-t') \right]  G(t') dt';\label{signal}
\end{equation}
The nuclear part of the response $R_n(t)$ is \cite{hellwarth_70}:
\begin{equation}
R_n(t)\propto -\frac{\partial}{\partial t}\langle\chi_{xy}(t)\chi_{xy}(0)\rangle
\label{signaleq}
\end{equation}
where $\chi_{xy}$ is the anisotropic component of the susceptibility tensor.
The frequency-dependent response can be obtained by a Fourier transform of $R_n(t)$ and can be related to the signal measured in a depolarized light scattering experiment \cite{bartolini_08,taschin_14}. 

A detailed description of the optical set-up and of the experimental apparatus is reported in reference \cite{taschin_14}. In summary: the laser system is a self-mode-locked Ti:sapphire laser producing pulses of 20 fs duration with energy of 3 nJ; the heterodyne-detection follows the configuration introduced in Ref.\cite{giraud_03}, where circularly polarized probe and differential acquisition of two opposite-phase signals on a balanced double photodiode is used, and the photodiode output signal is processed by a lock-in amplifier phase locked to the reference frequency at which the pump beam is chopped.

The extraction of the OKE response from the measured signal requires the knowledge of the instrumental function $G(t)$; this is a fundamental and not trivial experimental problem that becomes particularly critical for weak signals characterized by complex relaxation dynamics, as in the case of bulk water \cite{taschin_13,taschin_14} or confined water \cite{taschin_13b}. In this experiment the $G(t)$ was obtained by measuring the OKE signal of a CaF$_2$ plate of the same thickness of Vycor sample placed side by side in the same cell. The switching from the water measurement configuration to the instrumental one was achieved by simply translating the cell perpendicularly to the optical axis of the experiment without even touching the rest of optical set-up.

The HD-OKE experiments require a very good optical quality of the samples investigated. In spite of the good surface quality, the Vycor samples still produce a relatively high scattering of laser beams. The interference effects due to the scattered pump laser were removed by inserting a piezo-driven vibrating mirror in the optical path of the pump arm \cite{taschin_13b}.

\section{Sample}

We nanoconfined water in Vycor porous glasses (code 7930 by Corning Company). This is a well known porous glass having a solid structure and presenting pore diameters of 4 nm with low diameter dispersion. Other specifics can be found in the manufacturer technical data sheet \footnote{ http://www.corning.com/docs/specialtymaterials/pisheets/Vycor\%207930.pdf}. 
Our samples are Vycor slabs of 8x8x2 mm$^{3}$ dimensions. The samples were cleaned with 35\% hydrogen peroxide solution (heating up to 90 $^\circ$C for 2 hours) and washed in distilled water. They were then stored in P$_{2}$O$_{5}$ (phosphoric anhydride) until usage. 

Dry sample were obtained by heating Vycor at 400 $^\circ$C for 10 h, while samples at different hydration level have been prepared \textit{via} a vapor phase exothermic transfer from the bulk \cite{tombari_05}; all the samples were weighted for checking the water content and immediately closed into a hermetic cell for spectroscopic measurements. No significant variation of the sample mass was found after the OKE measurements. 
The ratio between the weight of water contained into Vycor glass and the weight of the borosilicate glass composing the Vycor matrix defines the sample hydration level. This is named \textit{Filling Fraction} parameter: $f$=H$_2$O (g) / Vycor (g). 

We performed HD-OKE experiment on 3 samples; a ``dry'' Vycor and two hydrated slabs. The first sample hydrated at $f$=11.6 $\pm$ 0.5 \% that corresponds to the presence of about ``two water layers'' on the pore surfaces and 50\% of permitted filling water; the second at $f$=24.3 $\pm$ 0.9 \%  that is the ``full hydration'' condition \cite{tombari_05}, i.e. 100\% of filling water. The water content was accurately determined from their weights and verified by FTIR spectrum.  

More details on the sample preparation and characterization by FTIR spectroscopy are reported in a previous paper \cite{taschin_13b}.

\section{Results on bulk water}

First of all, we measured the HD-OKE data of bulk water (i.e. not confined in Vycor) by the same experimental set-up and procedures. These data represent our benchmark of water dynamics that will enable us to perform a comparison with the HD-OKE data of water nanoconfined.

The HD-OKE data of liquid and supercooled bulk water provide unique information on the vibrational dynamics and structural relaxation taking place in water. The theoretical interpretation and the importance of HD-OKE data in the discussion of water anomalies has been reported in Ref.\cite{torre_04,taschin_13,taschin_14}.
\begin{figure} [htb]
\centering
\includegraphics[scale=.8]{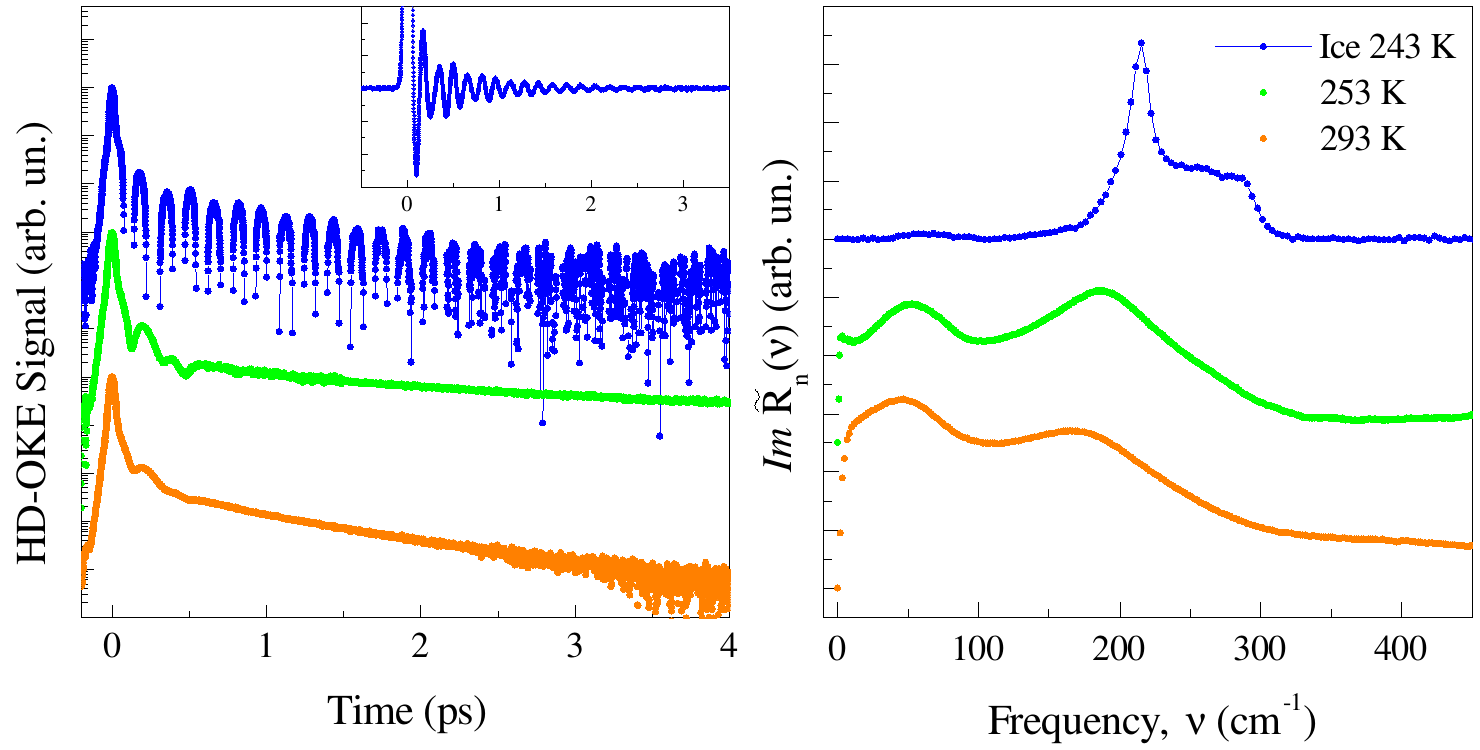}
\caption{We report the HD-OKE data of bulk water, vertically shifted for better visibility. In the left panel, we show the HD-OKE signals in a log-linear plot at 243 K, 253 K and 293 K temperatures (from top), in the inset there is the linear-linear plot of the lower temperature data corresponding to the hexagonal ice crystal phase. In the right panel, we show the Fourier transform of the HD-OKE data deconvoluted from the instrumental function (i.e the OKE response function in the frequency domain) \cite{taschin_13,taschin_14}.}
\label{bulk}
\end{figure} 
In figure \ref{bulk} we report HD-OKE data measured in bulk water in three different states, from the top: crystal, supercooled and liquid. In the left panel we report the data in the time-domain; in the right panel we show the OKE response function in the frequency-domain. The time-domain data of the liquid and supercooled phases clearly show the signature of fast oscillatory dynamics at short times, extending up to 1 ps, followed by a slower monotonic relaxation. The initial oscillatory component is due to the intermolecular vibrational dynamics and the slower decay to the structural relaxation dynamics \cite{torre_04,taschin_13,taschin_14}. The lower temperature data shows only an under-dumped oscillation without slower relaxation processes, this is the signature of ice crystal dynamics.

In order to get the spectra of the HD-OKE response function, we Fourier transform the measured data, we deconvolute them from the instrumental response and we retain the imaginary part of it: $Im[\tilde{R}_n(\nu)] \propto Im\lbrace FT\left[ S(t) \right] /FT\left[ G(t)\right] \rbrace$, see also eq.\ref{signal}. The details about these procedures are reported in Taschin et al. \cite{taschin_13,taschin_14}. The $Im[\tilde{R}_n(\nu)]$ obtained from the HD-OKE data are reported in right panel of figure \ref{bulk}. As expected, the liquid and the supercooled data present very similar dynamic features characterized by a shoulder appearing at very low frequency, $\nu \lesssim$ 10 cm$^{-1}$, due to the relaxation processes and by the two intermolecular vibrational bands around 50 and 175~cm$^{-1}$, named the ``bending" and ``stretching" modes, respectively. The ice crystal phase, hexagonal ice, shows a clearly different dynamics with an arrested structural and low frequency dynamics, the vibrational spectrum for $\nu>$150 cm$^{-1}$ presents the typical spectrum of ice \cite{kanno_98}.

\section{Results on nanoconfined water}

We measured the HD-OKE response on Vycor samples at different temperatures from ambient down to the crystallization occurrence for three different hydrations: dry, $f$= 11.6~\% (bilayer hydration) and 24.3~\% (full hydration) . 
\begin{figure} [htb]
\centering
\includegraphics[scale=0.8]{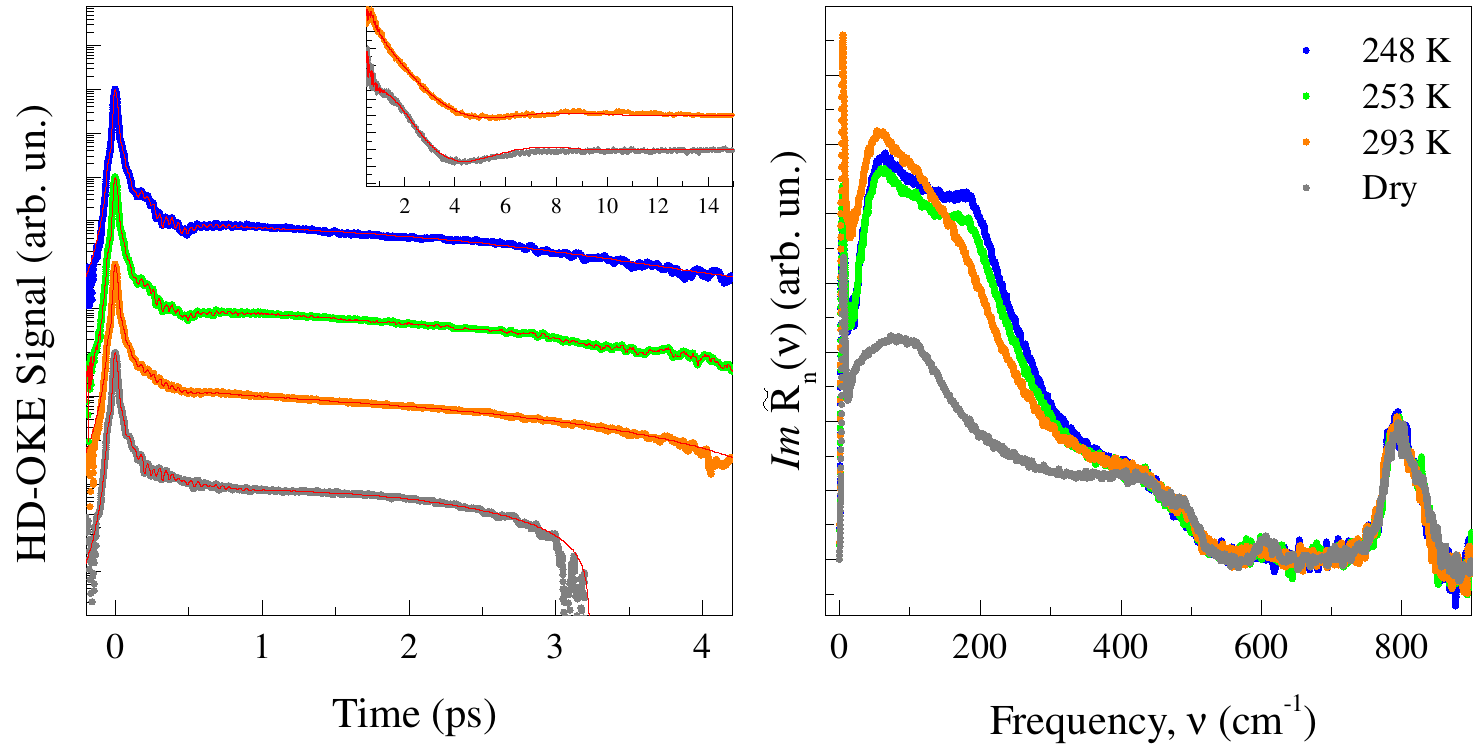}
\caption{We report HD-OKE data of nanoconfined water at $f$=11.6~\% corresponding to a bilayer water hydration at different temperatures together with the dry sample data. In the left panel the data are reported in a log-linear plot (Full circles: experimental data;  continuous line: best fit results). In the inset two signal decays (data on dry sample and on bilayer hydrated sample at 293 K) at longer times are shown in a linear-linear plot. The data have been vertically shifted to make all kinetics clearly visible. In the right panel: Fourier transforms of the HD-OKE data, deconvoluted from the instrumental response according to the procedure described in \cite{taschin_13b,taschin_14}. The data have been re-normalized on the high frequency peak, $\nu\approx$ 800 cm$^{-1}$, corresponding to the transverse optical mode of the silica matrix \cite{pilla_03}.}
\label{bilayer}
\end{figure}
\begin{figure} [htb]
\centering
\includegraphics[scale=0.8]{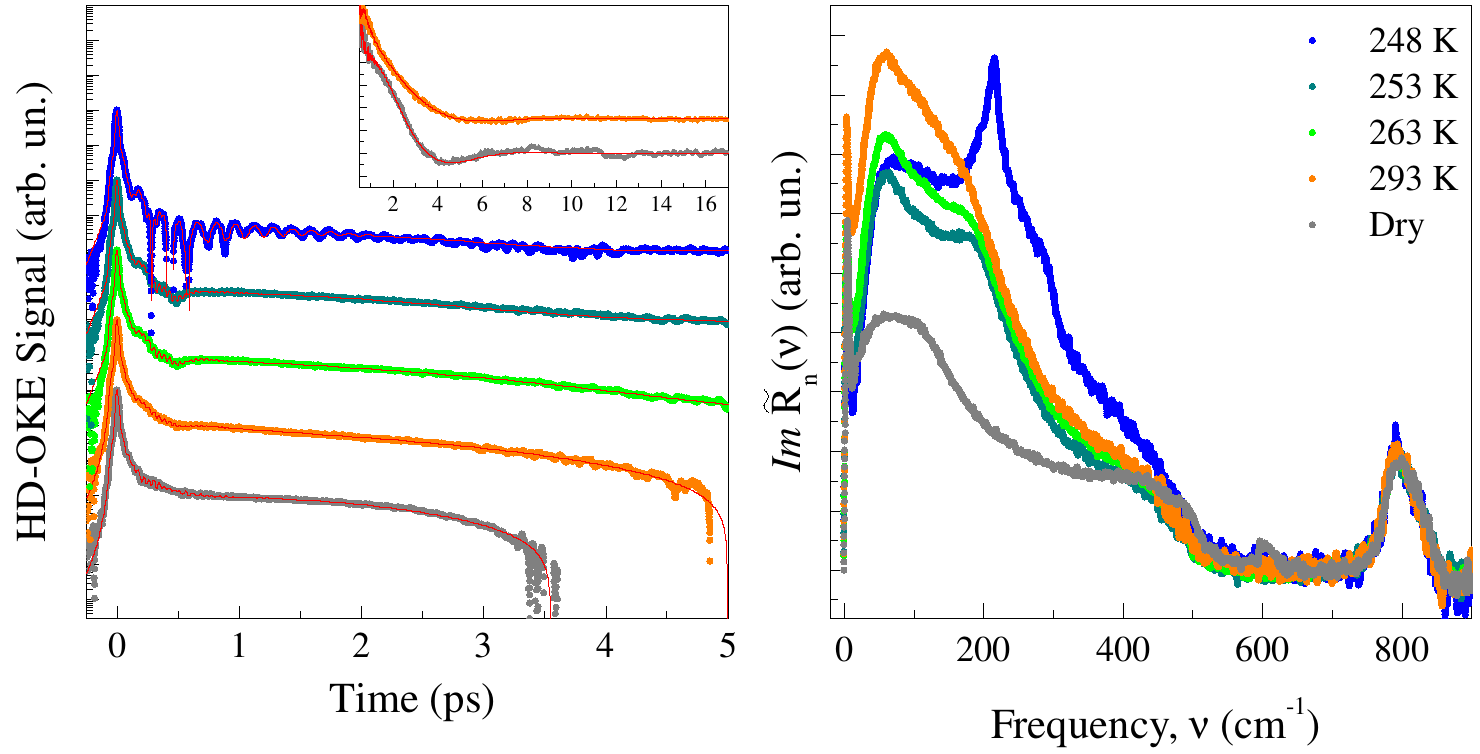}
\caption{HD-OKE data of nanoconfined water at full hydration $f$=24.3~\% with the best fit results (continuous lines). In the left panel the data are reported in a log-linear plot, while decays at longer times are shown in the inset. In the rigth panel we report the Fourier transforms of the HD-OKE data, deconvoluted from the instrumental response, on nanoconfined water at full hydration at different temperatures.}
\label{full}
\end{figure}

The experimental results with $f$= 11.6~\% hydration are collected in figure \ref{bilayer}, those for the $f$= 24.3~\% sample in figure \ref{full}. The left panels show in a log-linear plot the data in the time; the results for longer delay times are shown in the insets in a linear-linear plot. A relatively slow oscillation extending in the picosecond time-scale is present, see both insets of the left panels. This is likely due to an acoustic-like vibration localized on the pore surface \cite{taschin_13b,vacher_90}. In the hydrated Vycor the presence of liquid water adds a monotonic decay that becomes the dominant feature at full hydration; this contribution is attributed to the relaxation processes of nanoconfined water. In the right panels the Fourier transform of the HD-OKE response are reported. 

\begin{figure} [htb]
\centering
\includegraphics[scale=0.8]{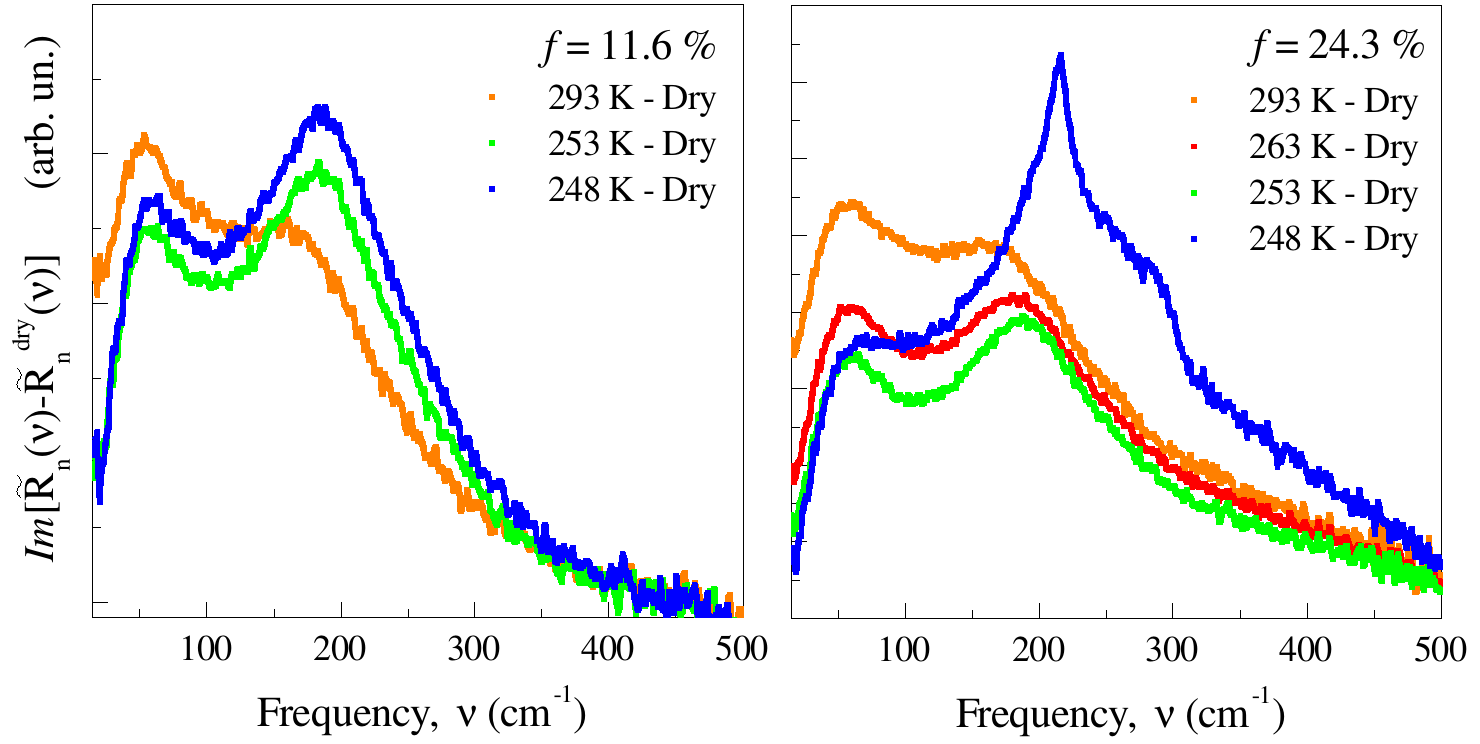}
\caption{Here we report the Fourier transforms of the HD-OKE data, deconvoluted from the instrumental response, on nanoconfined water after we subtracted the dry sample contribution. In the left panel the spectrum of bilayer water is reported, in the right that of fully hydrated Vycor sample.}
\label{difference}
\end{figure}

Just a simple look to these data reveals that there are not signature of ice formation at any temperature for the sample with $f$= 11.6~\%, whereas the presence of ice is clear in $f$= 24.3~\% sample at 248 K, the lower temperature measured. In fact this appears as a clear under-dumped oscillation (i.e. a sharp peak at about 215 cm$^{-1}$ in the frequency domain) easily detectable in the both the time and frequency domain data. 
The supercooled nanoconfined water shows a dynamics apparently similar to the bulk water; in fact it is characterized by a spectrum extending up to about 400 cm$^{-1}$ presenting two broad vibrational peaks very similar to the bending and stretching bands, see also figure \ref{difference}. A closer view shows how the positions and relative amplitudes are modified both in the bilayer and fully hydrated samples. As we mentioned, the lower frequency range is characterized by a low peak attributed to an acoustic-like pore vibration and a shoulder that is related to the structural relaxation process. The slow dynamics, which is hardly visible in the frequency-domain, it is better extracted from the time-domain data using a fitting procedure.
  
The rigorous definition of the OKE response function for water is a very complex issue \cite{taschin_14}, that becomes even more problematic for nanoconfined water. Here we decided to use a simple function in order to fit the data and provide an immediate comparison between the bulk and nanoconfined features. The response is taken as \cite{taschin_13b}:
\begin{equation}
R_n(t)=AR_{dry}(t)+B\frac{d}{dt} exp\left[-\left(\frac{t}{\tau}\right)^{\beta}\right]+\sum_i C_i exp\left(-\gamma_i^2 t^2\right)\sin\left(\omega_i t\right)
\label{response}	
\end{equation}
Eq.~\ref{signal} describes the convolution of the response function, $R_n(t)$, with the instrumental function, $G(t)$, where the $\delta$-function reproduces the instantaneous electronic response. Eq.~\ref{response} gives the response function simulating the material dynamics. The liquid water dynamics is described by the sum of  relaxation functions in the form of the time derivative of a stretched exponential \cite{torre_04}, and of a few damped oscillators (DO). The main fitting parameters of the model are: the structural relaxation time $\tau$, the stretching factor $\beta$,  the frequencies $\omega_i$ and the damping constants $\gamma_i$ of the DOs; the dry Vycor matrix response is taken into account adding a $R_{dry}(t)$ function, which is simulated by a sum of a series of damped oscillators and exponential functions, whose characteristic parameters were determined by fitting the HD-OKE signal of dry sample.

The equation \ref{response} can be used to fit both the bulk (with $A=0$) and the nanoconfined water (with $A \neq 0$) data; the results from the fits are reported as red continuous lines in the left panels of figures \ref{bilayer} and \ref{full}.

The fast part of the HD-OKE data has been fitted using a series of DOs (i.e. last term in equation \ref{response}) with frequency and damping parameters showing a smooth temperature dependence. Without enter in a detailed analysis of the fitting parameters the Fourier transform of the measured data enables a direct visualization of the frequency components present in the spectrum. We report in figure \ref{difference} the Fourier transforms of all the measured data after the spectrum of dry sample has been subtracted; this would isolate the signal contributions coming from the nanoconfined water.
The spectrum of nanoconfined water in both samples for T $>$ 248 K shows features similar to the bulk water; as expected these similarities are larger in the full hydration water than in the bilayer water \cite{taschin_13b}. The stretching and bending vibrational bands are clearly present and they show an increase of the intensity of the higher frequency component respect to the lower frequency part with decreasing temperatures. 
At the lowest measured temperature, T = 248 K, the bilayer water does not crystallize whereas the sample with full hydration does crystallize. The presence of ice inside the nanopores is proved by the evident and the relatively narrow peak appearing at the frequency of $\nu \approx$ 215 cm$^{-1}$, typical of external vibrations of ice crystals \cite{kanno_98}. This spectrum features can certainly be assigned to ice as proven by the measured data reported in figure \ref{bulk}. We can not distinguish between the cubic or hexagonal ice but, according to the previous reported neutron scattering data  \cite{bellissent_98,dore_00,bellissent_01}, it is very likely  cubic ice. 
 
The spectrum of nanoconfined water at 248 K, reported in the right panel of figure \ref{difference}, shows a broad band below the ice peaks that is not present in the bulk ice spectrum reported in right panel of figure \ref{bulk}. This broad vibrational component, extending over the enter frequency window reported, is due to those parts of water that remains mobile avoiding the freezing process. This water retains a vibrational dynamics and a finite structural relaxation time. The comparison of the reported experimental spectra suggests that the amount of water remaining in a supercooled state is a not negligible part of the nanoconfined water. We estimated from the spectrum area to be about 50\% of the total amount of internal water. This is in fairly agreement with the volumes occupied by the outer water (i.e. two water layers at the pore surface) and inner water (i.e. the remaining 4-5 layers of water  filling the internal part of the pore).

\begin{figure} [htb]
\centering
\includegraphics[scale=.6]{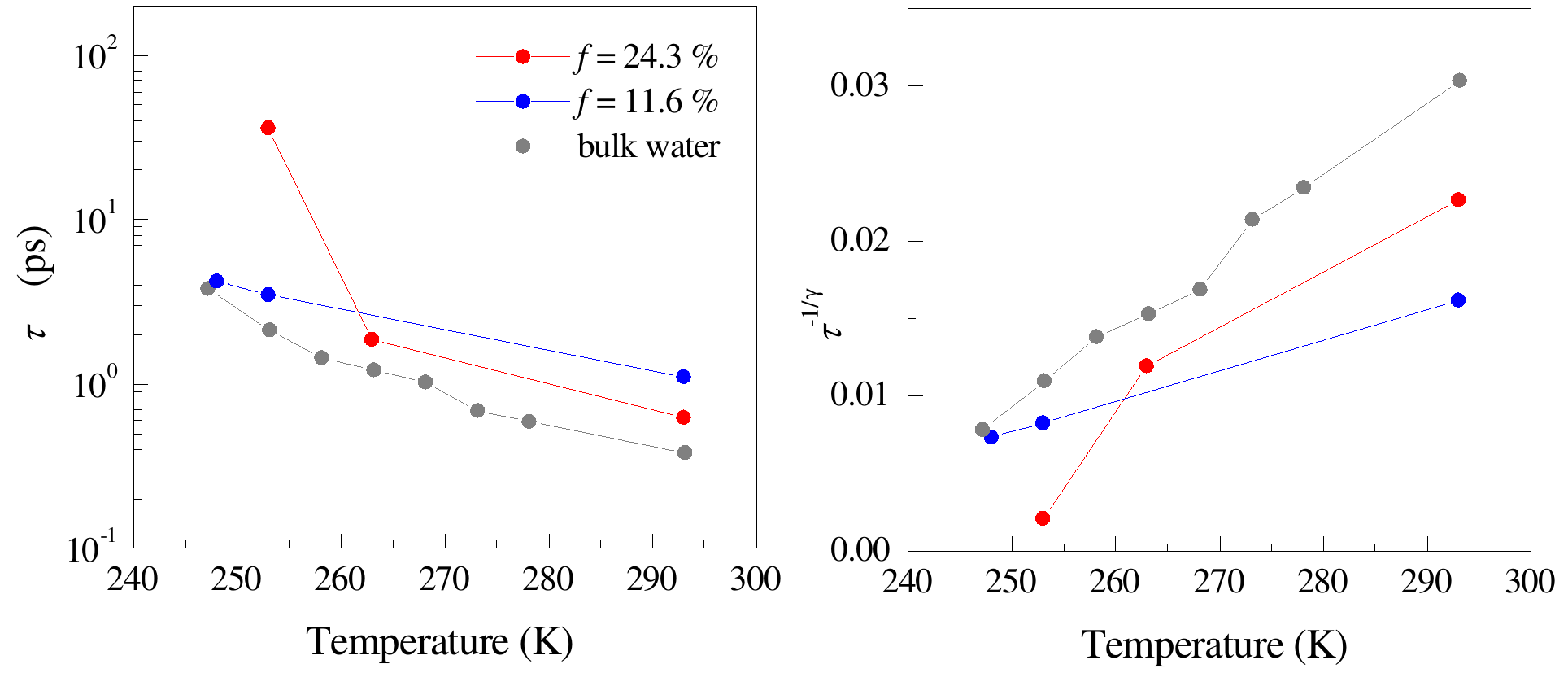}
\caption{In the left panel, we report the structural relaxation times obtained by the best fit procedure for the bilayer, $f$=11.6~\%, and fully hydrated, $f$=24.3~\%, samples; we report also the relaxation times obtained from the fits of the bulk water data \cite{taschin_13,taschin_14}. In the right panel we report the same data in a linearized plot to visualize the scaling proprieties, we used the critical exponent $\gamma=1.7$ found for bulk water.}
\label{RelaxTime}
\end{figure}

The best fit procedure enables to get the structural relaxation times of nanoconfined water, these are reported in figure \ref{RelaxTime}. The stretched exponential provides good fits at all the investigated temperatures but the fitting procedure does not enable to extract a reliable value for the stretching exponent; we fixed it to the value found for bulk water. Moreover the presence of the slow oscillating signal reduces the accuracy of the fitting parameters.   
The results reported in the figure~\ref{RelaxTime} show that the structural relaxation time does vary appreciably from bilayer hydration, $f$=11.6~\%, to full hydration, $f$=24.3~\%, both as value and temperature dependence. 

The structural relaxation time of nanoconfined water at lower hydration, i.e. two water layers, experiences a temperature dependence reduced respect to the bulk water that turns into an increase of the characteristic time scale of about a factor 4 at higher temperatures. This dynamics is strongly affected by the interactions with the silica surfaces; this was proven by the previous HD-OKE measurements with increasing hydrations \cite{taschin_13b} and it is confirmed by the present data of structural dynamics that shows a surprising weak temperature dependence.

The structural relaxation times at full hydration and high temperature are indeed intermediate between the bulk and bilayer water relaxation times. Moreover they show a slowing down process approaching the lower temperatures stronger than that taking place in bulk water; data suggest that reducing the temperature the molecular network is more stabilized in fully hydrated pores than in partially hydrated sample. At 248 K, when part the water is freeze, the HD-OKE still detects a very slow decay in the signal that suggests the presence of mobile water inside the pores. 

The temperature dependence of bulk water follows a power law dependence, $\tau\propto(T-T_c)^{-\gamma}$, with a critical exponent $\gamma=1.7$ and a critical temperature $T_c=227$ K \cite{taschin_13}. In the right panel of figure \ref{RelaxTime} we report the relaxation times in a linearized plot in order to verify if this power law apply also to nanoconfined water. Even if the measured temperatures are the minimum to verify a linear dependence, we find that both the hydration water possibly follows a power law with similar critical exponents but with different critical temperatures; lower for the bilayer water and higher for the full hydration water.

\section{Conclusion}

These results, as the previous studies, can be rationalized considering the presence of two water types inside the hydrophilic nanopores: the water close to pore surface, \textit{outer water}, and the more internal part of water, \textit{inner water}. We must distinguish the outer water (i.e. water layers in contact with the porous silica but at full hydration) from the \textit{multilayer water}: water layers at low hydration levels in contact with the porous silica and water vapour. Unfortunately, our experimental investigations cannot measure selectively the outer water dynamics, but only the multilayer water. In fact only the numerical simulation can probe the outer/inner water components. Nevertheless, the experimental data at different level of hydrations can give meaningful indications  \cite{taschin_13b}.

According to our results, water nanoconfined in 4 nm hydrophilic pores shows a vibrational dynamics that has general features similar to the bulk water even in the supercooled phase. This is basically true for both partially hydrated (i.e. bilayer water) and for fully hydrated samples (i.e. outer/inner water). At the lower temperature, 248 K, the vibrational dynamics of fully hydrated samples is modified reporting the contemporary presence of ice (inner water that freeze) and supercooled water (outer water that remains mobile). 

Instead the structural dynamics of nanoconfined water shows several differences from the bulk water, as well as structural relaxation of the bilayer water from the full hydration water. The temperature dependence of the structural relaxation times of the various samples has a complex scenario that is not of immediate understanding.
The relaxation times in bilayer water are slower then in bulk water and they show a weaker temperature dependence. This is certainly due to the interactions of water layers with the pore surfaces, that hinder the structural rearrangements and frustrate the nucleation processes. These effects are likely weakly dependent by the temperature. In the full hydrated pores lowering the temperature the structural relaxation undergoes a slowing down phenomenon stronger than in bulk water. These experimental evidences could suggest the following physical scenario; In the supercooled phase the water local structures are unstable and subject to two opposite effects: the surface interactions and the nucleation processes taking place in the inner part of the pores. These instabilities would generate slow fluctuations affecting the structural relaxation times.

\section*{Acknowledgements}
This work was supported by Regione Toscana,  prog. POR-CRO-FSE-UNIFI-26, by Ente Cassa di Risparmio Firenze, prog. 2012-0584 and by MIUR, prog. PRIN-2010ERFKXL-004.  We acknowledge M. De Pas, A. Montori and M. Giuntini for providing their continuous assistance in the electronic set-ups ; R. Ballerini and A. Hajeb for the mechanical realizations.

\section*{References}


\begin{thebibliography}{10}
\expandafter\ifx\csname url\endcsname\relax
  \def\url#1{{\tt #1}}\fi
\expandafter\ifx\csname urlprefix\endcsname\relax\def\urlprefix{URL }\fi
\providecommand{\eprint}[2][]{\url{#2}}

\bibitem{brovchenko_08}
Brovchenko I and Oleinikova A 2008 {\em Interfacial and Confined Water\/}
  (Elsevier)

\bibitem{jahnert_08}
J\"{a}hnert S, {Vaca Ch\'{a}vez} F, Schaumann G~E, Schreiber A, Sch\"{o}nhoff M
  and Findenegg G~H 2008 {\em Phys. Chem. Chem. Phys.\/} {\bf 10} 6039 

\bibitem{gallo_10}
Gallo P, Rovere M and Chen S~H 2010 {\em J. Phys. Chem. Lett.\/} {\bf 1} 729

\bibitem{molinero_10}
de~la Llave E, Molinero V and Scherlis D~A 2010 {\em J. Chem. Phys.\/} {\bf
  133} 034513

\bibitem{oguni_11}
Oguni M, Kanke Y, Nagoe A and Namba S 2011 {\em J. Phys. Chem. B\/} {\bf 115}
  14023

\bibitem{lecaer_11}
Caer S~L, Pin S, Esnouf S, Raffy Q, Renault J~P, Brubach J~B, Creff G and Roy P
  2011 {\em Phys. Chem. Chem. Phys.\/} {\bf 13} 17658

\bibitem{milischuk_11}
Milischuk A~a and Ladanyi B~M 2011 {\em J. Chem. Phys.\/} {\bf 135} 174709 

\bibitem{giovambattista_12}
Giovambattista N, Rossky P and Debenedetti P 2012 {\em Annu. Rev. Phys.
  Chem.\/} {\bf 63} 179

\bibitem{limmer_12}
Limmer D~T and Chandler D 2012 {\em J. Chem. Phys.\/} {\bf 137} 044509

\bibitem{milischuk_12}
Milischuk A~A, Krewald V and Ladanyi B~M 2012 {\em J. Chem. Phys.\/} {\bf 136}
  224704

\bibitem{bertrand_13}
Bertrand C~E, Zhang Y and Chen S~H 2013 {\em Phys. Chem. Chem. Phys.\/} {\bf
  15} 721

\bibitem{bruni_98}
Bruni F, Ricci M~A and Soper A~K 1998 {\em J. Chem. Phys.\/} {\bf 109}
  1478

\bibitem{bellissent_98}
Bellissent-Funel M 1998 {\em Journal of molecular liquids\/} {\bf 78} 19

\bibitem{dore_00}
Dore J 2000 {\em Chemical Physics\/} {\bf 258} 327

\bibitem{bellissent_01}
Bellissent-Funel M~C 2001 {\em Journal of Physics: Condensed Matter\/} {\bf 13}
  9165

\bibitem{tombari_05}
Tombari E, Salvetti G, Ferrari C and Johari G~P 2005 {\em Phys. Chem. Chem.
  Phys.\/} {\bf 7}(19) 3407

\bibitem{tombari_05b}
Tombari E, Salvetti G, Ferrari C and Johari G~P 2005 {\em The Journal of
  Chemical Physics\/} {\bf 122} 104712

\bibitem{cucini_07}
Cucini R, Taschin A, Ziparo C, Bartolini P and Torre R 2007 {\em The Europ.
  Physical Jour. : Spec. Topics\/} {\bf 141} 133

\bibitem{taschin_07}
Taschin A, Cucini R, Ziparo C and Torre R 2007 {\em Philos. Mag.\/} {\bf 87}
  715

\bibitem{taschin_08}
Taschin A, Cucini R, Bartolini P and Torre R 2008 {\em Europhys. Lett.\/} {\bf
  81} 58003

\bibitem{cucini_10}
Cucini R, Taschin A, Bartolini P and Torre R 2010 {\em J. Mech. Phys. Solids\/}
  {\bf 58} 1302

\bibitem{cucini_10b}
Cucini R, Taschin A, Bartolini P and Torre R 2010 {\em J. of Physics: Conf.
  Series\/} {\bf 214} 012032

\bibitem{taschin_10}
Taschin A, Cucini R, Bartolini P and Torre R 2010 {\em Europhys. Lett.\/} {\bf
  92} 26005

\bibitem{taschin_13b}
Taschin A, Bartolini P, Marcelli A, Righini R and Torre R 2013 {\em Faraday
  Discussions\/} {\bf 167} 293

\bibitem{hunt_07}
Hunt N~T, Jaye A~A and Meech S~R 2007 {\em Phys. Chem. Chem. Phys.\/} {\bf 9}
  2167

\bibitem{bartolini_08}
Bartolini P, Taschin A, Eramo R and Torre R 2008 {\em Optical Kerr Effect
  Experiments on Complex Liquids, A Direct Access to Fast Dynamic Processes\/}
  (New York: Springer) chap~2, pp 73--127

\bibitem{ricci_95}
Ricci M, Torre R, Foggi P, Kamalov V and Righini R 1995 {\em J. Chem. Phys.\/}
  {\bf 102} 9537

\bibitem{bartolini_99}
Bartolini P, Ricci M, Torre R, Righini R and Santa I 1999 {\em J. Chem.
  Phys.\/} {\bf 110} 8653

\bibitem{torre_95}
Torre R, Ricci M, Saielli G, Bartolini P and Righini R 1995 {\em Mol. Cryst.
  Liq. Cryst.\/} {\bf 262} 391

\bibitem{torre_98b}
Torre R, Tempestini F, Bartolini P and Righini R 1998 {\em Philos. Mag. B\/}
  {\bf 77} 645

\bibitem{ricci_04}
Ricci M, Wiebel S, Bartolini P, Taschin A and Torre R 2004 {\em Philos. Mag.\/}
  {\bf 84} 1491

\bibitem{torre_04}
Torre R, Bartolini P and Righini R 2004 {\em Nature\/} {\bf 428} 296

\bibitem{taschin_13}
Taschin A, Bartolini P, Eramo R, Righini R and Torre R 2013 {\em Nature
  Communications\/} {\bf 4} 2401

\bibitem{taschin_14}
Taschin A, Bartolini P, Eramo R, Righini R and Torre R 2014 {\em arxiv\/}
  1--15 (\textit{Preprint} \eprint{1406.5504})

\bibitem{torre_98}
Torre R, Bartolini P and Pick R 1998 {\em Phys. Rev. E\/} {\bf 57} 1912

\bibitem{torre_00}
Torre R, Bartolini P, Ricci M and Pick R 2000 {\em Europhys. Lett.\/} {\bf 52}
  324

\bibitem{ricci_02}
Ricci M, PBartolini and Torre R 2002 {\em Philos. Mag. B\/} {\bf 82} 541

\bibitem{prevosto_02}
Prevosto D, Bartolini P, Torre R, Ricci M, Taschin A, Capaccioli S, Lucchesi M
  and Rolla P 2002 {\em Phys. Rev. E\/} {\bf 66} 11502

\bibitem{pratesi_03}
Pratesi G, Bartolini P, Senatra D, Ricci M, Barocchi F, Righini R and Torre R
  2003 {\em Phys. Rev. E\/} {\bf 67} 021505

\bibitem{hellwarth_70}
Hellwarth R~W 1970 {\em J. Chem. Phys.\/} {\bf 52} 2128

\bibitem{giraud_03}
Giraud G, Gordon C, Dunkin I and Wynne K 2003 {\em J. Chem. Phys.\/} {\bf 119}
  464

\bibitem{kanno_98}
Kanno H, Tomikawa Y and Mishima O 1998 {\em Chem. Phys. Lett.\/} {\bf 293}
  412

\bibitem{pilla_03}
Pilla O, Fontana A, Caponi S, Rossi F, Viliani G, Gonzalez M, Fabiani E and
  Varsamis C 2003 {\em J. Non-Cryst. Solids\/} {\bf 322} 53

\bibitem{vacher_90}
Woignier T, Sauvajol J, Pelous J and Vacher R 1990 {\em J. Non-Cryst. Solids\/}
  {\bf 121} 206

\end{thebibliography}

\providecommand{\newblock}{}

\end{document}